%% file: main.tex
\documentclass[submission,copyright,creativecommons]{eptcs}
\input{preamble.tex}

\usepackage{tikzfig}
\usepackage{multirow}
\include{tikzstyle}
\usepackage{breakurl}
\providecommand{\event}{QPL 2014} 

\title{The ZX-calculus is incomplete for quantum mechanics}
\author{Christian Schr\"oder de Witt
\institute{Scherenbergstr. 22, 10439 Berlin}
\email{caschroeder@outlook.com}
\and
Vladimir Zamdzhiev
\institute{Department of Computer Science\\
Univeristy of Oxford\\
Oxford, United Kingdom}
\email{\quad vladimir.zamdzhiev@cs.ox.ac.uk}
}
\def\titlerunning{The ZX-calculus is incomplete for quantum mechanics}
\def\authorrunning{C. Schr\"oder de Witt \& V. Zamdzhiev}
\begin{document}
\maketitle

\begin{abstract}
We prove that the ZX-calculus is incomplete for quantum mechanics. We suggest the addition of a new 'color-swap' rule, of which currently no analytical formulation is known and which we suspect may be necessary, but not sufficient to make the ZX-calculus complete.
\end{abstract}

  \section{Introduction}
  \input{intro.tex}

  \section{Incompleteness}
  \input{incompleteness.tex}

  \section{Conclusion and Future Work}
  \input{conclusion.tex}

  \bibliographystyle{eptcs}
  \bibliography{References,bib}{}

\end{document}

%% file: preamble.tex
\usepackage{amsmath}
\usepackage{amssymb}
\usepackage{amsfonts}
\usepackage{amsthm}
\usepackage{enumerate}
\usepackage{amssymb}
\usepackage{textcomp}
\usepackage[english]{babel}
\usepackage[pdftex]{graphicx}
\usepackage{hyperref}
\usepackage{textcomp}
\usepackage[hypcap]{caption}
\usepackage[T1]{fontenc}
\usepackage[latin9]{inputenc}
\usepackage{graphicx}
\usepackage{babel}
\usepackage[usenames,dvipsnames]{color}
\usepackage{mathtools}
\usepackage{braket}
\usepackage{placeins}
\usepackage{color, colortbl}
\usepackage{fancyhdr}
\usepackage{float}
\usepackage{algorithm}
\usepackage{algpseudocode}
\usepackage{stmaryrd}

\numberwithin{equation}{section}

\theoremstyle{definition}

\theoremstyle{remark}

\date{\today}

\newcolumntype{C}{>{\centering\arraybackslash} m{3cm} }

\input{tikzstyle.tex}

%% file: tikzstyle.tex
\usepackage[svgnames]{xcolor}
\usepackage{tikz}
\usepackage{graphics}

\usetikzlibrary{decorations.pathmorphing}
\usetikzlibrary{decorations.markings}
\usetikzlibrary{decorations.pathreplacing}
\usetikzlibrary{arrows}
\usetikzlibrary{shapes.geometric}

\pgfdeclarelayer{edgelayer}
\pgfdeclarelayer{nodelayer}
\pgfsetlayers{edgelayer,nodelayer,main}

\tikzstyle{none}=[inner sep=0pt]
\tikzstyle{gn}=[circle,fill=Lime,draw=Black,line width=0.8 pt,inner sep=0pt,font=\bfseries]
\tikzstyle{rn}=[circle,fill=Red,draw=Black,line width=0.8 pt,inner sep=0pt,font=\bfseries,text=White]
\tikzstyle{H}=[rectangle,fill=Yellow,draw=Black]
\tikzstyle{line}=[scalar,fill=White,draw=Black]
\tikzstyle{io}=[rectangle,fill=White,draw=Black]
\tikzstyle{block}=[rectangle,fill=Orange,draw=Black]
\tikzstyle{graph}=[circle,fill=White,draw=Black]
\tikzstyle{empty}=[rectangle,fill=White,draw=White]
\tikzstyle{box}=[rectangle,fill=White,draw=Black]
\tikzstyle{dot}=[circle,fill=Black,draw=Black,inner sep=0pt,minimum size=1pt]
\tikzstyle{Dot}=[circle,fill=Black,draw=Black,inner sep=0pt,minimum size=3pt]
\tikzstyle{diam}=[rectangle,fill=Black,draw,yscale=1.2,rotate=45]
\tikzstyle{gangle}=[rectangle,fill=Lime,draw=Black]
\tikzstyle{rangle}=[rectangle,fill=Red,draw=Black]
\tikzstyle{circ}=[circle,fill=none,draw=Black,scale=1.3]
\tikzstyle{bbox}=[rectangle,fill=Blue,draw=Blue,scale=0.6]

\tikzstyle{simple}=[-,draw=Black]
\tikzstyle{directed}=[->,draw=Black]
\tikzstyle{blue}=[-,draw=Blue]
\tikzstyle{blued}=[->,draw=Blue]

\tikzstyle{every picture}=[baseline=-0.25em]


%% file: intro.tex
Coecke and Abramsky pioneered the field of categorical quantum mechanics in
\cite{cqm}. Later, from this study, an intuitive graphical calculus (dubbed the
ZX-calculus) was developed by Coecke and Duncan \cite{bigmainzx}\cite{zx_small}, which can be used
as an alternative to Dirac notation in a wide number of
applications \cite{strongcompl}\cite{hillebrand_qpl12}\cite{quanto_steane_code}\cite{zx_mbqc}.

Backens recently proved that the ZX-calculus is complete for an important
subset of quantum mechanics, namely stabilizer quantum mechanics, i.e. that for
stabilizer quantum mechanics, any equation that can be shown to hold in the
Dirac formalism can also be shown to hold within the
ZX-calculus\cite{miriam_complete_sqm}. For her proof, she relied on
operations on a special class of quantum states, namely graph states.  This
paper adresses the question of whether the ZX-calculus is complete for the
whole of quantum mechanics, and the answer is found to be negative.

\subsection{Syntax and Semantics of the ZX-calculus}

The syntax and semantics of the ZX-calculus are presented below.
The semantics are given in Hilbert space. We begin with atomic diagrams.
The inputs to the diagrams are located at the bottom and the outputs are
located at the top. 

\begin{align*}
\left \llbracket \quad
\input{tikz/id.tikz}
\quad \right \rrbracket
 = 
 \begin{pmatrix}
  1 & 0  \\
  0 & 1  
 \end{pmatrix}
 = I
\mbox \quad \quad \quad
\left \llbracket \quad
\input{tikz/swap.tikz} 
\quad \right \rrbracket
 = 
 \begin{pmatrix}
  1 & 0 & 0 & 0 \\
  0 & 0 & 1 & 0\\
  0 & 1 & 0 & 0 \\
  0 & 0 & 0 & 1
 \end{pmatrix}
=\sigma
\end{align*}

\begin{align*}
\left \llbracket \quad
\begin{aligned}
\input{tikz/cup.tikz} 
\end{aligned}
\quad \right \rrbracket
 = 
 \bra{00} + \bra{11}
\quad\quad
\left \llbracket \quad
\begin{aligned}
\input{tikz/cap.tikz} 
\end{aligned}
\quad \right \rrbracket
 = 
 \begin{aligned}
 \ket{00} + \ket{11}
\end{aligned}
\left \llbracket \quad
\begin{aligned}
\input{tikz/h.tikz} 
\end{aligned}
\quad \right \rrbracket
 = 
 \begin{aligned}
 \frac 1 {\sqrt{2}}
 \begin{pmatrix}
  1 & 1  \\
  1 & -1  
 \end{pmatrix}
 = H
\end{aligned}
\end{align*}
\begin{align*}
\left \llbracket \quad
\begin{aligned}
\input{tikz/green_sem.tikz} 
\end{aligned}
\quad \right \rrbracket
 = 
 \begin{aligned}
 \begin{cases}
  \ket{0^m} &\mapsto \ket{0^n}\\
  \ket{1^m} &\mapsto e^{i\alpha} \ket{1^n}\\
  \text{rest} &\mapsto 0
 \end{cases}
\end{aligned}
\left \llbracket \quad
\begin{aligned}
\input{tikz/red_sem.tikz} 
\end{aligned}
\quad \right \rrbracket
 = 
 \begin{aligned}
 \begin{cases}
  \ket{+^m} &\mapsto \ket{+^n}\\
  \ket{-^m} &\mapsto e^{i\alpha} \ket{-^n}\\
  \text{rest} &\mapsto 0
 \end{cases}
\end{aligned}
\end{align*}

where in the last two diagrams $m$ is the number of inputs and $n$ is the
number of outputs. The labels of the red and green dots form the circle group
under addition. So, admissible values are $\alpha \in [0, 2\pi).$ We also make
the convention that we will not write a label for the points when $\alpha=0.$

We can create compound diagrams from smaller diagrams in two ways - either placing two diagrams next to each horizontally, or plugging the outputs of one diagram to 
the inputs of another. If

$$
\left \llbracket \quad
\begin{aligned}
\input{tikz/d1.tikz} 
\end{aligned}
\quad \right \rrbracket
 = 
 \begin{aligned}
 D_1 \text{\quad\quad\quad and \quad\quad}
  \end{aligned}
\left \llbracket \quad
  \begin{aligned}
\input{tikz/d2.tikz} 
\end{aligned}
\quad \right \rrbracket
 = 
 \begin{aligned}
 D_2
  \end{aligned}
$$

then

$$
\left \llbracket \quad
\begin{aligned}
\input{tikz/comp1.tikz} 
\end{aligned}
\quad \right \rrbracket
 = 
 \begin{aligned}
 D_1 \otimes D_2
 \end{aligned}
$$
and
$$
\left \llbracket \quad
\begin{aligned}
\input{tikz/comp2.tikz} 
\end{aligned}
\quad \right \rrbracket
 = 
 \begin{aligned}
 D_1 \circ D_2
 \end{aligned}
$$
In the latter diagram, the number of outputs of $\Psi_2$ has to be the same as
the number of inputs of $\Psi_1.$

By following the above rules we can represent any pure state map on qubits as a
diagram in the ZX-calculus \cite{bigmainzx}.

\subsection{ZX Equational Rules}

The rules of the ZX-calculus are given by:\\\quad\\
\input{includes/zx_rules.tex}

\subsection{Completeness, Soundness and Universality}

The original rules of the ZX-calculus, as put forward by Coecke and Duncan
\cite{bigmainzx} did not contain the Euler decomposition of the Hadamard
gate, \textbf{EU}. Subsequently, Duncan and Perdrix
proved \cite{euler_necessity} that the rule \textbf{EU} was not derivable from
within the original ZX-calculus. The original ZX-calculus was therefore
\textit{incomplete}. Informally, incompleteness signifies that there are
equations that can be proven to hold in Dirac-von Neumann notation that cannot
be proven in the ZX-calculus. This would reduce the power of the graphical
calculus and possibly limit its applications in automated reasoning.

Backens showed in \cite{miriam_complete_sqm} that the current ZX-calculus,
which is simply the original ZX-calculus extended by \textbf{EU}, is
complete for an important fragment of quantum mechanics, namely
\textit{stabilizer quantum mechanics (SQM)}. Her proof relies on the fact
that each SQM state is, under local Clifford
operations \cite{gottesman_heisenberg_1999}, equivalent to a special entangled
state, namely a \textit{graph state} \cite{hein_multiparty_2004}. This allows
one to abstract away from matrix representations and instead decide equivalence
between different SQM states by performing \textit{local complementations}, a
class of graph manipulations, between graph states. In this way, Backens showed
that SQM states may be represented by so-called rGS-LC diagrams, which are only
equivalent iff they are graphically identical. In this way, equivalence can be
decided in the ZX-calculus.

However, ideally, one would wish the ZX-calculus to be as physically expressive
as the complete Dirac-von Neumann formalism. To this goal, three important
properties of the calculus need to be established: universality, soundness and
completeness.

The ZX-calculus is \textit{sound} \cite{bigmainzx}. That is, if
$ZX \vdash D_1 = D_2$ then 
$\llbracket D_1 \rrbracket = e^{i \phi} \llbracket D_2 \rrbracket$. 
In other words, if two diagrams are equal under the axioms of the ZX-calculus,
then their Hilbert space interpretations are equal up to a global
phase.

Secondly, the ZX-calculus is \textit{universal}, meaning that it can express
any quantum state and gate. This is easily proven by showing that the
ZX-calculus can express any of the set of universal quantum
gates \cite{bigmainzx}. 

Finally, \textit{completeness} is the converse of soundness. That is, if  
$\llbracket D_1 \rrbracket = \llbracket D_2 \rrbracket$ then
$ZX \vdash D_1 = D_2$. In the next section, we will show that the ZX-calculus
does not have this property.

%% file: tikz/id.tikz
\begin{tikzpicture}
	\begin{pgfonlayer}{nodelayer}
		\node [style=none] (0) at (0, 1) {};
		\node [style=none] (1) at (0, -0.75) {};
	\end{pgfonlayer}
	\begin{pgfonlayer}{edgelayer}
		\draw (0) to (1);
	\end{pgfonlayer}
\end{tikzpicture}

%% file: tikz/swap.tikz
\begin{tikzpicture}
	\begin{pgfonlayer}{nodelayer}
		\node [style=none] (0) at (-1, 1) {};
		\node [style=none] (1) at (0.75, 1) {};
		\node [style=none] (2) at (-1, -1) {};
		\node [style=none] (3) at (0.75, -1) {};
	\end{pgfonlayer}
	\begin{pgfonlayer}{edgelayer}
		\draw (3) to (0);
		\draw (2) to (1);
	\end{pgfonlayer}
\end{tikzpicture}

%% file: tikz/cup.tikz
\begin{tikzpicture}
	\begin{pgfonlayer}{nodelayer}
		\node [style=none] (0) at (-0.5, -0.75) {};
		\node [style=none] (1) at (0.75, -0.75) {};
	\end{pgfonlayer}
	\begin{pgfonlayer}{edgelayer}
		\draw [bend left=90, looseness=2.00] (0) to (1);
	\end{pgfonlayer}
\end{tikzpicture}

%% file: tikz/cap.tikz
\begin{tikzpicture}
	\begin{pgfonlayer}{nodelayer}
		\node [style=none] (0) at (-1.5, 0.5) {};
		\node [style=none] (1) at (-0.25, 0.5) {};
	\end{pgfonlayer}
	\begin{pgfonlayer}{edgelayer}
		\draw [bend left=270, looseness=1.75] (0) to (1);
	\end{pgfonlayer}
\end{tikzpicture}

%% file: tikz/h.tikz
\begin{tikzpicture}
	\begin{pgfonlayer}{nodelayer}
		\node [style=none] (0) at (0, 1) {};
		\node [style=H] (1) at (0, 0) {H};
		\node [style=none] (2) at (0, -1) {};
	\end{pgfonlayer}
	\begin{pgfonlayer}{edgelayer}
		\draw (0) to (1);
		\draw (1) to (2);
	\end{pgfonlayer}
\end{tikzpicture}

%% file: tikz/green_sem.tikz
\begin{tikzpicture}
	\begin{pgfonlayer}{nodelayer}
		\node [style=none] (0) at (-2.75, 0.75) {};
		\node [style=none] (1) at (-1.75, 0.75) {};
		\node [style=dot, scale=0.3] (2) at (-1, 0.75) {};
		\node [style=dot, scale=0.3] (3) at (-0.5, 0.75) {};
		\node [style=dot, scale=0.3] (4) at (0, 0.75) {};
		\node [style=none] (5) at (0.75, 0.75) {};
		\node [style=gn] (6) at (-1, -0.25) {$\alpha$};
		\node [style=none] (7) at (-2.75, -1.25) {};
		\node [style=none] (8) at (-1.75, -1.25) {};
		\node [style=dot, scale=0.3] (9) at (-1, -1.25) {};
		\node [style=dot, scale=0.3] (10) at (-0.5, -1.25) {};
		\node [style=dot, scale=0.3] (11) at (0, -1.25) {};
		\node [style=none] (12) at (0.75, -1.25) {};
	\end{pgfonlayer}
	\begin{pgfonlayer}{edgelayer}
		\draw [bend left] (7) to (6);
		\draw [bend left] (8) to (6);
		\draw [bend left=15] (5) to (6);
		\draw [bend right] (1) to (6);
		\draw [bend right, looseness=1.25] (0) to (6);
		\draw [bend left=15] (6) to (12);
	\end{pgfonlayer}
\end{tikzpicture}

%% file: tikz/red_sem.tikz
\begin{tikzpicture}
	\begin{pgfonlayer}{nodelayer}
		\node [style=none] (0) at (-2.75, 0.75) {};
		\node [style=none] (1) at (-1.75, 0.75) {};
		\node [style=dot, scale=0.3] (2) at (-1, 0.75) {};
		\node [style=dot, scale=0.3] (3) at (-0.5, 0.75) {};
		\node [style=dot, scale=0.3] (4) at (0, 0.75) {};
		\node [style=none] (5) at (0.75, 0.75) {};
		\node [style=rn] (6) at (-1, -0.25) {$\alpha$};
		\node [style=none] (7) at (-2.75, -1.25) {};
		\node [style=none] (8) at (-1.75, -1.25) {};
		\node [style=dot, scale=0.3] (9) at (-1, -1.25) {};
		\node [style=dot, scale=0.3] (10) at (-0.5, -1.25) {};
		\node [style=dot, scale=0.3] (11) at (0, -1.25) {};
		\node [style=none] (12) at (0.75, -1.25) {};
	\end{pgfonlayer}
	\begin{pgfonlayer}{edgelayer}
		\draw [bend left] (7) to (6);
		\draw [bend right, looseness=1.25] (0) to (6);
		\draw [bend left=15] (5) to (6);
		\draw [bend right] (1) to (6);
		\draw [bend left=15] (6) to (12);
		\draw [bend left] (8) to (6);
	\end{pgfonlayer}
\end{tikzpicture}

%% file: tikz/d1.tikz
\begin{tikzpicture}
	\begin{pgfonlayer}{nodelayer}
		\node [style=none] (0) at (-2.75, 0.75) {};
		\node [style=none] (1) at (-1.75, 0.75) {};
		\node [style=dot, scale=0.3] (2) at (-1, 0.75) {};
		\node [style=dot, scale=0.3] (3) at (-0.5, 0.75) {};
		\node [style=dot, scale=0.3] (4) at (0, 0.75) {};
		\node [style=none] (5) at (0.75, 0.75) {};
		\node [style=block] (6) at (-1, -0.25) {$\Psi_1$};
		\node [style=none] (7) at (-2.75, -1.25) {};
		\node [style=none] (8) at (-1.75, -1.25) {};
		\node [style=dot, scale=0.3] (9) at (-1, -1.25) {};
		\node [style=dot, scale=0.3] (10) at (-0.5, -1.25) {};
		\node [style=dot, scale=0.3] (11) at (0, -1.25) {};
		\node [style=none] (12) at (0.75, -1.25) {};
	\end{pgfonlayer}
	\begin{pgfonlayer}{edgelayer}
		\draw [bend left=15] (6) to (12);
		\draw [bend left=15] (5) to (6);
		\draw [bend right, looseness=1.25] (0) to (6);
		\draw [bend left] (7) to (6);
		\draw [bend right] (1) to (6);
		\draw [bend left] (8) to (6);
	\end{pgfonlayer}
\end{tikzpicture}

%% file: tikz/d2.tikz
\begin{tikzpicture}
	\begin{pgfonlayer}{nodelayer}
		\node [style=none] (0) at (-2.75, 0.75) {};
		\node [style=none] (1) at (-1.75, 0.75) {};
		\node [style=dot, scale=0.3] (2) at (-1, 0.75) {};
		\node [style=dot, scale=0.3] (3) at (-0.5, 0.75) {};
		\node [style=dot, scale=0.3] (4) at (0, 0.75) {};
		\node [style=none] (5) at (0.75, 0.75) {};
		\node [style=block] (6) at (-1, -0.25) {$\Psi_2$};
		\node [style=none] (7) at (-2.75, -1.25) {};
		\node [style=none] (8) at (-1.75, -1.25) {};
		\node [style=dot, scale=0.3] (9) at (-1, -1.25) {};
		\node [style=dot, scale=0.3] (10) at (-0.5, -1.25) {};
		\node [style=dot, scale=0.3] (11) at (0, -1.25) {};
		\node [style=none] (12) at (0.75, -1.25) {};
	\end{pgfonlayer}
	\begin{pgfonlayer}{edgelayer}
		\draw [bend left=15] (6) to (12);
		\draw [bend left] (8) to (6);
		\draw [bend right, looseness=1.25] (0) to (6);
		\draw [bend left=15] (5) to (6);
		\draw [bend right] (1) to (6);
		\draw [bend left] (7) to (6);
	\end{pgfonlayer}
\end{tikzpicture}

%% file: tikz/comp1.tikz
\begin{tikzpicture}
	\begin{pgfonlayer}{nodelayer}
		\node [style=none] (0) at (-2.75, 0.75) {};
		\node [style=none] (1) at (-1.75, 0.75) {};
		\node [style=dot, scale=0.3] (2) at (-1, 0.75) {};
		\node [style=dot, scale=0.3] (3) at (-0.5, 0.75) {};
		\node [style=dot, scale=0.3] (4) at (0, 0.75) {};
		\node [style=none] (5) at (0.75, 0.75) {};
		\node [style=none] (6) at (1.75, 0.75) {};
		\node [style=none] (7) at (2.75, 0.75) {};
		\node [style=dot, scale=0.3] (8) at (3.5, 0.75) {};
		\node [style=dot, scale=0.3] (9) at (4, 0.75) {};
		\node [style=dot, scale=0.3] (10) at (4.5, 0.75) {};
		\node [style=none] (11) at (5.25, 0.75) {};
		\node [style=block] (12) at (-1, -0.25) {$\Psi_1$};
		\node [style=block] (13) at (3.5, -0.25) {$\Psi_2$};
		\node [style=none] (14) at (-2.75, -1.25) {};
		\node [style=none] (15) at (-1.75, -1.25) {};
		\node [style=dot, scale=0.3] (16) at (-1, -1.25) {};
		\node [style=dot, scale=0.3] (17) at (-0.5, -1.25) {};
		\node [style=dot, scale=0.3] (18) at (0, -1.25) {};
		\node [style=none] (19) at (0.75, -1.25) {};
		\node [style=none] (20) at (1.75, -1.25) {};
		\node [style=none] (21) at (2.75, -1.25) {};
		\node [style=dot, scale=0.3] (22) at (3.5, -1.25) {};
		\node [style=dot, scale=0.3] (23) at (4, -1.25) {};
		\node [style=dot, scale=0.3] (24) at (4.5, -1.25) {};
		\node [style=none] (25) at (5.25, -1.25) {};
	\end{pgfonlayer}
	\begin{pgfonlayer}{edgelayer}
		\draw [bend left] (14) to (12);
		\draw [bend left=15] (5) to (12);
		\draw [bend left] (15) to (12);
		\draw [bend left] (20) to (13);
		\draw [bend left] (21) to (13);
		\draw [bend right] (1) to (12);
		\draw [bend right] (7) to (13);
		\draw [bend left=15] (13) to (25);
		\draw [bend right, looseness=1.25] (0) to (12);
		\draw [bend right, looseness=1.25] (6) to (13);
		\draw [bend left=15] (12) to (19);
		\draw [bend left=15] (11) to (13);
	\end{pgfonlayer}
\end{tikzpicture}

%% file: tikz/comp2.tikz
\begin{tikzpicture}
	\begin{pgfonlayer}{nodelayer}
		\node [style=none] (0) at (1.75, 3.5) {};
		\node [style=none] (1) at (2.75, 3.5) {};
		\node [style=dot, scale=0.3] (2) at (3.5, 3.5) {};
		\node [style=dot, scale=0.3] (3) at (4, 3.5) {};
		\node [style=dot, scale=0.3] (4) at (4.5, 3.5) {};
		\node [style=none] (5) at (5.25, 3.5) {};
		\node [style=block] (6) at (3.5, 2.5) {$\Psi_1$};
		\node [style=dot, scale=0.3] (7) at (3.5, 1.25) {};
		\node [style=dot, scale=0.3] (8) at (4, 1.25) {};
		\node [style=dot, scale=0.3] (9) at (4.5, 1.25) {};
		\node [style=block] (10) at (3.5, -0.25) {$\Psi_2$};
		\node [style=none] (11) at (1.75, -1.25) {};
		\node [style=none] (12) at (2.75, -1.25) {};
		\node [style=dot, scale=0.3] (13) at (3.5, -1.25) {};
		\node [style=dot, scale=0.3] (14) at (4, -1.25) {};
		\node [style=dot, scale=0.3] (15) at (4.5, -1.25) {};
		\node [style=none] (16) at (5.25, -1.25) {};
	\end{pgfonlayer}
	\begin{pgfonlayer}{edgelayer}
		\draw [bend left] (12) to (10);
		\draw [bend right] (1) to (6);
		\draw [bend left=15] (5) to (6);
		\draw [bend left=15] (10) to (16);
		\draw [bend right=75, looseness=3.00] (10) to (6);
		\draw [bend right, looseness=1.25] (0) to (6);
		\draw [bend left=45, looseness=1.50] (10) to (6);
		\draw [bend left=90, looseness=2.00] (10) to (6);
		\draw [bend left] (11) to (10);
	\end{pgfonlayer}
\end{tikzpicture}

%% file: includes/zx_rules.tex
\begin{minipage}{1.0\textwidth}
\begin{center}
\begin{tabular}{ccl}
\multicolumn{2}{c}{\textit{"Only the topology matters"}} &  \textbf{(T)}\\[3em]
$%
\InputIfFileExists{tikz/ZXpt_S1greenL.tikz}{}{\input{./figures/tikz/ZXpt_S1greenL.tikz}}
 = %
\InputIfFileExists{tikz/ZXpt_S1greenR.tikz}{}{\input{./figures/tikz/ZXpt_S1greenR.tikz}}
$ & $%
\InputIfFileExists{tikz/ZXpt_S1redL.tikz}{}{\input{./figures/tikz/ZXpt_S1redL.tikz}}
 = %
\InputIfFileExists{tikz/ZXpt_S1redR.tikz}{}{\input{./figures/tikz/ZXpt_S1redR.tikz}}
$ & \textbf{(S1)}\\[5em]
\multirow{2}{*}{$%
\InputIfFileExists{tikz/ZXpt_S2_1_1.tikz}{}{\input{./figures/tikz/ZXpt_S2_1_1.tikz}}
 = %
\InputIfFileExists{tikz/ZXpt_S2_1_2.tikz}{}{\input{./figures/tikz/ZXpt_S2_1_2.tikz}}
 = %
\InputIfFileExists{tikz/ZXpt_S2_1_3.tikz}{}{\input{./figures/tikz/ZXpt_S2_1_3.tikz}}
$} & $%
\InputIfFileExists{tikz/ZXpt_S2_2u_1.tikz}{}{\input{./figures/tikz/ZXpt_S2_2u_1.tikz}}
 = %
\InputIfFileExists{tikz/ZXpt_S2_2u_2.tikz}{}{\input{./figures/tikz/ZXpt_S2_2u_2.tikz}}
 = %
\InputIfFileExists{tikz/ZXpt_S2_2u_3.tikz}{}{\input{./figures/tikz/ZXpt_S2_2u_3.tikz}}
$ & \textbf{(S2)} \\[1em]
& $%
\InputIfFileExists{tikz/ZXpt_S2_2l_1.tikz}{}{\input{./figures/tikz/ZXpt_S2_2l_1.tikz}}
 = %
\InputIfFileExists{tikz/ZXpt_S2_2l_2.tikz}{}{\input{./figures/tikz/ZXpt_S2_2l_2.tikz}}
 = %
\InputIfFileExists{tikz/ZXpt_S2_2l_3.tikz}{}{\input{./figures/tikz/ZXpt_S2_2l_3.tikz}}
$ & \quad \\[3em]
$%
\InputIfFileExists{tikz/ZXpt_B1_1L.tikz}{}{\input{./figures/tikz/ZXpt_B1_1L.tikz}}
 = %
\InputIfFileExists{tikz/ZXpt_B1_1R.tikz}{}{\input{./figures/tikz/ZXpt_B1_1R.tikz}}
$ \quad $%
\InputIfFileExists{tikz/ZXpt_B1_2L.tikz}{}{\input{./figures/tikz/ZXpt_B1_2L.tikz}}
 = %
\InputIfFileExists{tikz/ZXpt_B1_2R.tikz}{}{\input{./figures/tikz/ZXpt_B1_2R.tikz}}
$ \textbf{(B1)} & $%
\InputIfFileExists{tikz/ZXpt_B2_L.tikz}{}{\input{./figures/tikz/ZXpt_B2_L.tikz}}
 = %
\InputIfFileExists{tikz/ZXpt_B2_R.tikz}{}{\input{./figures/tikz/ZXpt_B2_R.tikz}}
$ & \textbf{(B2)}\\[3em]
$%
\InputIfFileExists{tikz/ZXpt_K1_L1.tikz}{}{\input{./figures/tikz/ZXpt_K1_L1.tikz}}
 = %
\InputIfFileExists{tikz/ZXpt_K1_R1.tikz}{}{\input{./figures/tikz/ZXpt_K1_R1.tikz}}
$ \quad $%
\InputIfFileExists{tikz/ZXpt_K1_L2.tikz}{}{\input{./figures/tikz/ZXpt_K1_L2.tikz}}
 = %
\InputIfFileExists{tikz/ZXpt_K1_R2.tikz}{}{\input{./figures/tikz/ZXpt_K1_R2.tikz}}
$ \textbf{(K1)} & $%
\InputIfFileExists{tikz/ZXpt_K2_L1.tikz}{}{\input{./figures/tikz/ZXpt_K2_L1.tikz}}
 = %
\InputIfFileExists{tikz/ZXpt_K2_R1.tikz}{}{\input{./figures/tikz/ZXpt_K2_R1.tikz}}
$ \quad $%
\InputIfFileExists{tikz/ZXpt_K2_L2.tikz}{}{\input{./figures/tikz/ZXpt_K2_L2.tikz}}
 = %
\InputIfFileExists{tikz/ZXpt_K2_R2.tikz}{}{\input{./figures/tikz/ZXpt_K2_R2.tikz}}
$ & \textbf{(K2)}\\[3em]
$%
\InputIfFileExists{tikz/ZXpt_C_L.tikz}{}{\input{./figures/tikz/ZXpt_C_L.tikz}}
 = %
\InputIfFileExists{tikz/ZXpt_C_R.tikz}{}{\input{./figures/tikz/ZXpt_C_R.tikz}}
$ \textbf{(C)} & $%
\InputIfFileExists{tikz/ZXpt_euler_L.tikz}{}{\input{./figures/tikz/ZXpt_euler_L.tikz}}
 = %
\InputIfFileExists{tikz/ZXpt_euler_R.tikz}{}{\input{./figures/tikz/ZXpt_euler_R.tikz}}
$ & \textbf{(EU)} \\[5em]
$%
\InputIfFileExists{tikz/ZXpt_D1_L.tikz}{}{\input{./figures/tikz/ZXpt_D1_L.tikz}}
 = %
\InputIfFileExists{tikz/ZXpt_D1_R.tikz}{}{\input{./figures/tikz/ZXpt_D1_R.tikz}}
$ \textbf{(D1)}& $%
\InputIfFileExists{tikz/ZXpt_D2_L.tikz}{}{\input{./figures/tikz/ZXpt_D2_L.tikz}}
 = %
\InputIfFileExists{tikz/ZXpt_D2_R.tikz}{}{\input{./figures/tikz/ZXpt_D2_R.tikz}}
$  & \textbf{(D2)}\\[3em]
\end{tabular}
\end{center}
\end{minipage}

%% file: tikz/ZXpt_S1greenL.tikz
\begin{tikzpicture}
	\begin{pgfonlayer}{nodelayer}
		\node [style=gn] (0) at (-0.5000003, 0.2500001) {$\alpha$};
		\node [style=gn] (1) at (0.5000003, -0.2500001) {$\beta$};
		\node [style=none] (2) at (-0.9999998, 1.25) {};
		\node [style=none] (3) at (0, 1.25) {};
		\node [style=none] (4) at (-0.5000003, -1.25) {};
		\node [style=none] (5) at (-1.5, -1.25) {};
		\node [style=none] (6) at (0, -1.25) {};
		\node [style=none] (7) at (0.9999998, -1.25) {};
		\node [style=none] (8) at (0.5000003, 1.25) {};
		\node [style=none] (9) at (1.5, 1.25) {};
		\node [style=none] (10) at (0, -0) {...};
		\node [style=none] (11) at (-0.9999998, -1.25) {...};
		\node [style=none] (12) at (0.9999998, 1.25) {...};
		\node [style=none] (13) at (0.5000003, -1.25) {...};
		\node [style=none] (14) at (-0.5000003, 1.25) {...};
	\end{pgfonlayer}
	\begin{pgfonlayer}{edgelayer}
		\draw [bend left, looseness=0.75] (5.center) to (0);
		\draw (4.center) to (0);
		\draw [bend left, looseness=1.00] (0) to (1);
		\draw [bend right, looseness=1.00] (0) to (1);
		\draw [bend right=15, looseness=1.00] (2.center) to (0);
		\draw [bend right=15, looseness=1.00] (0) to (3.center);
		\draw (1) to (8.center);
		\draw [bend right, looseness=1.00] (1) to (9.center);
		\draw [bend right=15, looseness=1.00] (1) to (6.center);
		\draw [bend left=15, looseness=1.00] (1) to (7.center);
	\end{pgfonlayer}
\end{tikzpicture}

%% file: tikz/ZXpt_S1greenR.tikz
\begin{tikzpicture}
	\begin{pgfonlayer}{nodelayer}
		\node [style=none] (0) at (-0.9999998, 1.25) {};
		\node [style=none] (1) at (-0.5000003, 1.25) {};
		\node [style=none] (2) at (0.5000003, 1.25) {};
		\node [style=none] (3) at (0.9999998, 1.25) {};
		\node [style=none] (4) at (-0.9999998, -1.25) {};
		\node [style=none] (5) at (0.5000003, -1.25) {};
		\node [style=none] (6) at (0.9999998, -1.25) {};
		\node [style=none] (7) at (-0.5000003, -1.25) {};
		\node [style=gn] (8) at (0, -0) {$\alpha+\beta$};
		\node [style=none] (9) at (-0.7500001, 1.25) {...};
		\node [style=none] (10) at (0.7500001, 1.25) {...};
		\node [style=none] (11) at (-0.7500001, -1.25) {...};
		\node [style=none] (12) at (0.7500001, -1.25) {...};
	\end{pgfonlayer}
	\begin{pgfonlayer}{edgelayer}
		\draw [bend right, looseness=1.00] (0.center) to (8);
		\draw [bend right=15, looseness=1.00] (1.center) to (8);
		\draw [bend right=15, looseness=1.00] (8) to (2.center);
		\draw [bend right, looseness=1.00] (8) to (3.center);
		\draw [bend right, looseness=1.00] (8) to (4.center);
		\draw [bend right=15, looseness=1.00] (8) to (7.center);
		\draw [bend left=15, looseness=1.00] (8) to (5.center);
		\draw [bend left, looseness=1.00] (8) to (6.center);
	\end{pgfonlayer}
\end{tikzpicture}

%% file: tikz/ZXpt_S1redL.tikz
\begin{tikzpicture}
	\begin{pgfonlayer}{nodelayer}
		\node [style=rn] (0) at (-0.5000003, 0.2500001) {$\alpha$};
		\node [style=rn] (1) at (0.5000003, -0.2500001) {$\beta$};
		\node [style=none] (2) at (-0.9999998, 1.25) {};
		\node [style=none] (3) at (0, 1.25) {};
		\node [style=none] (4) at (-0.5000003, -1.25) {};
		\node [style=none] (5) at (-1.5, -1.25) {};
		\node [style=none] (6) at (0, -1.25) {};
		\node [style=none] (7) at (0.9999998, -1.25) {};
		\node [style=none] (8) at (0.5000003, 1.25) {};
		\node [style=none] (9) at (1.5, 1.25) {};
		\node [style=none] (10) at (0, -0) {...};
		\node [style=none] (11) at (-0.9999998, -1.25) {...};
		\node [style=none] (12) at (0.9999998, 1.25) {...};
		\node [style=none] (13) at (0.5000003, -1.25) {...};
		\node [style=none] (14) at (-0.5000003, 1.25) {...};
	\end{pgfonlayer}
	\begin{pgfonlayer}{edgelayer}
		\draw [bend left, looseness=0.75] (5.center) to (0);
		\draw (4.center) to (0);
		\draw [bend left, looseness=1.00] (0) to (1);
		\draw [bend right, looseness=1.00] (0) to (1);
		\draw [bend right=15, looseness=1.00] (2.center) to (0);
		\draw [bend right=15, looseness=1.00] (0) to (3.center);
		\draw (1) to (8.center);
		\draw [bend right, looseness=1.00] (1) to (9.center);
		\draw [bend right=15, looseness=1.00] (1) to (6.center);
		\draw [bend left=15, looseness=1.00] (1) to (7.center);
	\end{pgfonlayer}
\end{tikzpicture}

%% file: tikz/ZXpt_S1redR.tikz
\begin{tikzpicture}
	\begin{pgfonlayer}{nodelayer}
		\node [style=none] (0) at (-0.9999998, 1.25) {};
		\node [style=none] (1) at (-0.5000003, 1.25) {};
		\node [style=none] (2) at (0.5000003, 1.25) {};
		\node [style=none] (3) at (0.9999998, 1.25) {};
		\node [style=none] (4) at (-0.9999998, -1.25) {};
		\node [style=none] (5) at (0.5000003, -1.25) {};
		\node [style=none] (6) at (0.9999998, -1.25) {};
		\node [style=none] (7) at (-0.5000003, -1.25) {};
		\node [style=rn] (8) at (0, -0) {$\alpha+\beta$};
		\node [style=none] (9) at (-0.7500001, 1.25) {...};
		\node [style=none] (10) at (0.7500001, 1.25) {...};
		\node [style=none] (11) at (-0.7500001, -1.25) {...};
		\node [style=none] (12) at (0.7500001, -1.25) {...};
	\end{pgfonlayer}
	\begin{pgfonlayer}{edgelayer}
		\draw [bend right, looseness=1.00] (0.center) to (8);
		\draw [bend right=15, looseness=1.00] (1.center) to (8);
		\draw [bend right=15, looseness=1.00] (8) to (2.center);
		\draw [bend right, looseness=1.00] (8) to (3.center);
		\draw [bend right, looseness=1.00] (8) to (4.center);
		\draw [bend right=15, looseness=1.00] (8) to (7.center);
		\draw [bend left=15, looseness=1.00] (8) to (5.center);
		\draw [bend left, looseness=1.00] (8) to (6.center);
	\end{pgfonlayer}
\end{tikzpicture}

%% file: tikz/ZXpt_S2_1_1.tikz
\begin{tikzpicture}
	\begin{pgfonlayer}{nodelayer}
		\node [style=gn] (0) at (0, -0) {$\quad$};
		\node [style=none] (1) at (0, -0.75) {};
		\node [style=none] (2) at (0, 0.75) {};
	\end{pgfonlayer}
	\begin{pgfonlayer}{edgelayer}
		\draw (0) to (2.center);
		\draw (0) to (1.center);
	\end{pgfonlayer}
\end{tikzpicture}

%% file: tikz/ZXpt_S2_1_2.tikz
\begin{tikzpicture}
	\begin{pgfonlayer}{nodelayer}
		\node [style=none] (0) at (0, -0.75) {};
		\node [style=none] (1) at (0, 0.75) {};
	\end{pgfonlayer}
	\begin{pgfonlayer}{edgelayer}
		\draw (0.center) to (1.center);
	\end{pgfonlayer}
\end{tikzpicture}

%% file: tikz/ZXpt_S2_1_3.tikz
\begin{tikzpicture}
	\begin{pgfonlayer}{nodelayer}
		\node [style=rn] (0) at (0, -0) {$\quad$};
		\node [style=none] (1) at (0, -0.75) {};
		\node [style=none] (2) at (0, 0.75) {};
	\end{pgfonlayer}
	\begin{pgfonlayer}{edgelayer}
		\draw (0) to (2.center);
		\draw (0) to (1.center);
	\end{pgfonlayer}
\end{tikzpicture}

%% file: tikz/ZXpt_S2_2u_1.tikz
\begin{tikzpicture}
	\begin{pgfonlayer}{nodelayer}
		\node [style=gn] (0) at (0, 0.25) {$\quad$};
		\node [style=none] (1) at (-0.75, -0.25) {};
		\node [style=none] (2) at (0.75, -0.25) {};
	\end{pgfonlayer}
	\begin{pgfonlayer}{edgelayer}
		\draw [bend left, looseness=1.00] (1.center) to (0);
		\draw [bend left, looseness=1.00] (0) to (2.center);
	\end{pgfonlayer}
\end{tikzpicture}

%% file: tikz/ZXpt_S2_2u_2.tikz
\begin{tikzpicture}
	\begin{pgfonlayer}{nodelayer}
		\node [style=none] (0) at (0, 0.25) {};
		\node [style=none] (1) at (-0.75, -0.25) {};
		\node [style=none] (2) at (0.75, -0.25) {};
	\end{pgfonlayer}
	\begin{pgfonlayer}{edgelayer}
		\draw [bend left, looseness=1.00] (1.center) to (0.center);
		\draw [bend left, looseness=1.00] (0.center) to (2.center);
	\end{pgfonlayer}
\end{tikzpicture}

%% file: tikz/ZXpt_S2_2u_3.tikz
\begin{tikzpicture}
	\begin{pgfonlayer}{nodelayer}
		\node [style=rn] (0) at (0, 0.25) {$\quad$};
		\node [style=none] (1) at (-0.75, -0.25) {};
		\node [style=none] (2) at (0.75, -0.25) {};
	\end{pgfonlayer}
	\begin{pgfonlayer}{edgelayer}
		\draw [bend left, looseness=1.00] (1.center) to (0);
		\draw [bend left, looseness=1.00] (0) to (2.center);
	\end{pgfonlayer}
\end{tikzpicture}

%% file: tikz/ZXpt_S2_2l_1.tikz
\begin{tikzpicture}
	\begin{pgfonlayer}{nodelayer}
		\node [style=gn] (0) at (0, -0.25) {$\quad$};
		\node [style=none] (1) at (0.75, 0.25) {};
		\node [style=none] (2) at (-0.75, 0.25) {};
	\end{pgfonlayer}
	\begin{pgfonlayer}{edgelayer}
		\draw [bend left, looseness=1.00] (1.center) to (0);
		\draw [bend left, looseness=1.00] (0) to (2.center);
	\end{pgfonlayer}
\end{tikzpicture}

%% file: tikz/ZXpt_S2_2l_2.tikz
\begin{tikzpicture}
	\begin{pgfonlayer}{nodelayer}
		\node [style=none] (0) at (0, -0.25) {};
		\node [style=none] (1) at (0.75, 0.25) {};
		\node [style=none] (2) at (-0.75, 0.25) {};
	\end{pgfonlayer}
	\begin{pgfonlayer}{edgelayer}
		\draw [bend left, looseness=1.00] (1.center) to (0.center);
		\draw [bend left, looseness=1.00] (0.center) to (2.center);
	\end{pgfonlayer}
\end{tikzpicture}

%% file: tikz/ZXpt_S2_2l_3.tikz
\begin{tikzpicture}
	\begin{pgfonlayer}{nodelayer}
		\node [style=rn] (0) at (0, -0.25) {$\quad$};
		\node [style=none] (1) at (0.75, 0.25) {};
		\node [style=none] (2) at (-0.75, 0.25) {};
	\end{pgfonlayer}
	\begin{pgfonlayer}{edgelayer}
		\draw [bend left, looseness=1.00] (1.center) to (0);
		\draw [bend left, looseness=1.00] (0) to (2.center);
	\end{pgfonlayer}
\end{tikzpicture}

%% file: tikz/ZXpt_B1_1L.tikz
\begin{tikzpicture}
	\begin{pgfonlayer}{nodelayer}
		\node [style=diam] (0) at (-0.75, -0) {};
		\node [style=rn] (1) at (0.25, 0.5) {$\quad$};
		\node [style=gn] (2) at (0.25, -0.25) {$\quad$};
		\node [style=none] (3) at (-0.25, -0.75) {};
		\node [style=none] (4) at (0.75, -0.75) {};
	\end{pgfonlayer}
	\begin{pgfonlayer}{edgelayer}
		\draw [bend left, looseness=1.00] (3.center) to (2);
		\draw (2) to (1);
		\draw [bend left, looseness=1.00] (2) to (4.center);
	\end{pgfonlayer}
\end{tikzpicture}

%% file: tikz/ZXpt_B1_1R.tikz
\begin{tikzpicture}
	\begin{pgfonlayer}{nodelayer}
		\node [style=rn] (0) at (-0.25, 0.25) {$\quad$};
		\node [style=rn] (1) at (0.25, 0.25) {$\quad$};
		\node [style=none] (2) at (-0.25, -0.25) {};
		\node [style=none] (3) at (0.25, -0.25) {};
	\end{pgfonlayer}
	\begin{pgfonlayer}{edgelayer}
		\draw (2.center) to (0);
		\draw (3.center) to (1);
	\end{pgfonlayer}
\end{tikzpicture}

%% file: tikz/ZXpt_B1_2L.tikz
\begin{tikzpicture}
	\begin{pgfonlayer}{nodelayer}
		\node [style=diam] (0) at (-0.75, -0) {};
		\node [style=gn] (1) at (0.25, 0.5) {$\quad$};
		\node [style=rn] (2) at (0.25, -0.25) {$\quad$};
		\node [style=none] (3) at (-0.25, -0.75) {};
		\node [style=none] (4) at (0.75, -0.75) {};
	\end{pgfonlayer}
	\begin{pgfonlayer}{edgelayer}
		\draw [bend left, looseness=1.00] (3.center) to (2);
		\draw (2) to (1);
		\draw [bend left, looseness=1.00] (2) to (4.center);
	\end{pgfonlayer}
\end{tikzpicture}

%% file: tikz/ZXpt_B1_2R.tikz
\begin{tikzpicture}
	\begin{pgfonlayer}{nodelayer}
		\node [style=gn] (0) at (-0.25, 0.25) {$\quad$};
		\node [style=gn] (1) at (0.25, 0.25) {$\quad$};
		\node [style=none] (2) at (-0.25, -0.25) {};
		\node [style=none] (3) at (0.25, -0.25) {};
	\end{pgfonlayer}
	\begin{pgfonlayer}{edgelayer}
		\draw (2.center) to (0);
		\draw (3.center) to (1);
	\end{pgfonlayer}
\end{tikzpicture}

%% file: tikz/ZXpt_B2_L.tikz
\begin{tikzpicture}
	\begin{pgfonlayer}{nodelayer}
		\node [style=gn] (0) at (-0.25, 0.5000001) {$\quad$};
		\node [style=gn] (1) at (0.7500001, 0.5000001) {$\quad$};
		\node [style=rn] (2) at (-0.25, -0.5000001) {$\quad$};
		\node [style=rn] (3) at (0.7500001, -0.5000001) {$\quad$};
		\node [style=none] (4) at (-0.25, -1) {};
		\node [style=none] (5) at (0.7500001, -1) {};
		\node [style=none] (6) at (-0.25, 1) {};
		\node [style=none] (7) at (0.7500001, 1) {};
		\node [style=diam] (8) at (-1, -0) {};
	\end{pgfonlayer}
	\begin{pgfonlayer}{edgelayer}
		\draw (2) to (1);
		\draw (0) to (3);
		\draw [bend right, looseness=1.00] (0) to (2);
		\draw [bend left, looseness=1.00] (1) to (3);
		\draw (0) to (6.center);
		\draw (1) to (7.center);
		\draw (2) to (4.center);
		\draw (3) to (5.center);
	\end{pgfonlayer}
\end{tikzpicture}

%% file: tikz/ZXpt_B2_R.tikz
\begin{tikzpicture}
	\begin{pgfonlayer}{nodelayer}
		\node [style=rn] (0) at (0, 0.5000001) {$\quad$};
		\node [style=gn] (1) at (0, -0.5000001) {$\quad$};
		\node [style=none] (2) at (-0.5000001, -1) {};
		\node [style=none] (3) at (0.5000001, -1) {};
		\node [style=none] (4) at (-0.5000001, 1) {};
		\node [style=none] (5) at (0.5000001, 1) {};
	\end{pgfonlayer}
	\begin{pgfonlayer}{edgelayer}
		\draw [bend right=45, looseness=1.00] (4.center) to (0);
		\draw [bend right=45, looseness=1.00] (0) to (5.center);
		\draw (0) to (1);
		\draw [bend right=45, looseness=1.00] (1) to (2.center);
		\draw [bend left=45, looseness=1.25] (1) to (3.center);
	\end{pgfonlayer}
\end{tikzpicture}

%% file: tikz/ZXpt_K1_L1.tikz
\begin{tikzpicture}
	\begin{pgfonlayer}{nodelayer}
		\node [style=gn] (0) at (0, 0.25) {$\pi$};
		\node [style=rn] (1) at (0, -0.25) {$\quad$};
		\node [style=none] (2) at (-0.5000001, -0.7500001) {};
		\node [style=none] (3) at (0.5000001, -0.7500001) {};
		\node [style=none] (4) at (0, 0.75) {};
		\node [style=none] (5) at (0, -0.7500001) {...};
	\end{pgfonlayer}
	\begin{pgfonlayer}{edgelayer}
		\draw [bend left, looseness=1.00] (2.center) to (1);
		\draw [bend left, looseness=1.00] (1) to (3.center);
		\draw (1) to (0);
		\draw (0) to (4.center);
	\end{pgfonlayer}
\end{tikzpicture}

%% file: tikz/ZXpt_K1_R1.tikz
\begin{tikzpicture}
	\begin{pgfonlayer}{nodelayer}
		\node [style=rn] (0) at (0, 0.5000001) {$\quad$};
		\node [style=gn] (1) at (-0.5000001, -0.25) {$\pi$};
		\node [style=gn] (2) at (0.5000001, -0.25) {$\pi$};
		\node [style=none] (3) at (-0.5000001, -0.7500001) {};
		\node [style=none] (4) at (0.5000001, -0.7500001) {};
		\node [style=none] (5) at (0, 1) {};
		\node [style=none] (6) at (0, -0.7500001) {...};
	\end{pgfonlayer}
	\begin{pgfonlayer}{edgelayer}
		\draw (1) to (0);
		\draw (0) to (5.center);
		\draw (0) to (2);
		\draw (2) to (4.center);
		\draw (1) to (3.center);
	\end{pgfonlayer}
\end{tikzpicture}

%% file: tikz/ZXpt_K1_L2.tikz
\begin{tikzpicture}
	\begin{pgfonlayer}{nodelayer}
		\node [style=rn] (0) at (0, 0.25) {$\pi$};
		\node [style=gn] (1) at (0, -0.25) {$\quad$};
		\node [style=none] (2) at (-0.5000001, -0.7500001) {};
		\node [style=none] (3) at (0.5000001, -0.7500001) {};
		\node [style=none] (4) at (0, 0.75) {};
		\node [style=none] (5) at (0, -0.7500001) {...};
	\end{pgfonlayer}
	\begin{pgfonlayer}{edgelayer}
		\draw [bend left, looseness=1.00] (2.center) to (1);
		\draw [bend left, looseness=1.00] (1) to (3.center);
		\draw (1) to (0);
		\draw (0) to (4.center);
	\end{pgfonlayer}
\end{tikzpicture}

%% file: tikz/ZXpt_K1_R2.tikz
\begin{tikzpicture}
	\begin{pgfonlayer}{nodelayer}
		\node [style=gn] (0) at (0, 0.5000001) {$\quad$};
		\node [style=rn] (1) at (-0.5000001, -0.25) {$\pi$};
		\node [style=rn] (2) at (0.5000001, -0.25) {$\pi$};
		\node [style=none] (3) at (-0.5000001, -0.7500001) {};
		\node [style=none] (4) at (0.5000001, -0.7500001) {};
		\node [style=none] (5) at (0, 1) {};
		\node [style=none] (6) at (0, -0.7500001) {...};
	\end{pgfonlayer}
	\begin{pgfonlayer}{edgelayer}
		\draw (1) to (0);
		\draw (0) to (5.center);
		\draw (0) to (2);
		\draw (2) to (4.center);
		\draw (1) to (3.center);
	\end{pgfonlayer}
\end{tikzpicture}

%% file: tikz/ZXpt_K2_L1.tikz
\begin{tikzpicture}
	\begin{pgfonlayer}{nodelayer}
		\node [style=gn] (0) at (0, 0.25) {$\mbox{ } \pi \mbox{ }$};
		\node [style=rn] (1) at (0, -0.5) {$\mbox{ } \alpha \mbox{ }$};
		\node [style=none] (2) at (0, -1) {};
		\node [style=none] (3) at (0, 0.75) {};
	\end{pgfonlayer}
	\begin{pgfonlayer}{edgelayer}
		\draw (3.center) to (0);
		\draw (0) to (1);
		\draw (1) to (2.center);
	\end{pgfonlayer}
\end{tikzpicture}

%% file: tikz/ZXpt_K2_R1.tikz
\begin{tikzpicture}
	\begin{pgfonlayer}{nodelayer}
		\node [style=rn] (0) at (0, 0.25) {$-\alpha$};
		\node [style=gn] (1) at (0, -0.5) {$\mbox{ } \pi \mbox{ }$};
		\node [style=none] (2) at (0, -1) {};
		\node [style=none] (3) at (0, 0.75) {};
	\end{pgfonlayer}
	\begin{pgfonlayer}{edgelayer}
		\draw (3.center) to (0);
		\draw (0) to (1);
		\draw (1) to (2.center);
	\end{pgfonlayer}
\end{tikzpicture}

%% file: tikz/ZXpt_K2_L2.tikz
\begin{tikzpicture}
	\begin{pgfonlayer}{nodelayer}
		\node [style=rn] (0) at (0, 0.25) {$\mbox{ } \pi \mbox{ }$};
		\node [style=gn] (1) at (0, -0.5) {$\mbox{ } \alpha \mbox{ }$};
		\node [style=none] (2) at (0, -1) {};
		\node [style=none] (3) at (0, 0.75) {};
	\end{pgfonlayer}
	\begin{pgfonlayer}{edgelayer}
		\draw (3.center) to (0);
		\draw (0) to (1);
		\draw (1) to (2.center);
	\end{pgfonlayer}
\end{tikzpicture}

%% file: tikz/ZXpt_K2_R2.tikz
\begin{tikzpicture}
	\begin{pgfonlayer}{nodelayer}
		\node [style=gn] (0) at (0, 0.25) {$-\alpha$};
		\node [style=rn] (1) at (0, -0.5) {$\mbox{ } \pi \mbox{ }$};
		\node [style=none] (2) at (0, -1) {};
		\node [style=none] (3) at (0, 0.75) {};
	\end{pgfonlayer}
	\begin{pgfonlayer}{edgelayer}
		\draw (3.center) to (0);
		\draw (0) to (1);
		\draw (1) to (2.center);
	\end{pgfonlayer}
\end{tikzpicture}

%% file: tikz/ZXpt_C_L.tikz
\begin{tikzpicture}
	\begin{pgfonlayer}{nodelayer}
		\node [style=rn] (0) at (0, -0) {$\alpha$};
		\node [style=none] (1) at (-0.5000001, 1) {};
		\node [style=none] (2) at (0.5000001, 1) {};
		\node [style=none] (3) at (1.25, 1) {};
		\node [style=none] (4) at (-1.25, 1) {};
		\node [style=none] (5) at (-1.25, -0.9999999) {};
		\node [style=none] (6) at (-0.5000001, -0.9999999) {};
		\node [style=none] (7) at (0.5000001, -0.9999999) {};
		\node [style=none] (8) at (1.25, -0.9999999) {};
		\node [style=none] (9) at (0, 1) {...};
		\node [style=none] (10) at (0, -0.9999999) {...};
	\end{pgfonlayer}
	\begin{pgfonlayer}{edgelayer}
		\draw [bend right, looseness=1.00] (4.center) to (0);
		\draw [bend left=15, looseness=1.00] (0) to (1.center);
		\draw [bend right=15, looseness=1.00] (0) to (2.center);
		\draw [bend right, looseness=1.00] (0) to (3.center);
		\draw [bend left=15, looseness=1.00] (0) to (7.center);
		\draw [bend right=15, looseness=1.00] (0) to (6.center);
		\draw [bend right, looseness=1.00] (0) to (5.center);
		\draw [bend left, looseness=1.00] (0) to (8.center);
	\end{pgfonlayer}
\end{tikzpicture}

%% file: tikz/ZXpt_C_R.tikz
\begin{tikzpicture}
	\begin{pgfonlayer}{nodelayer}
		\node [style=gn] (0) at (0, -0) {$\alpha$};
		\node [style=none] (1) at (-0.5000001, 1.25) {};
		\node [style=none] (2) at (0.5000001, 1.25) {};
		\node [style=none] (3) at (1.5, 1.25) {};
		\node [style=none] (4) at (-1.5, 1.25) {};
		\node [style=none] (5) at (-1.5, -1.25) {};
		\node [style=none] (6) at (-0.5000001, -1.25) {};
		\node [style=none] (7) at (0.5000001, -1.25) {};
		\node [style=none] (8) at (1.5, -1.25) {};
		\node [style=none] (9) at (0, 1.25) {...};
		\node [style=none] (10) at (0, -1.25) {...};
		\node [style=H] (11) at (-1.5, 0.75) {H};
		\node [style=H] (12) at (-0.5000001, 0.75) {H};
		\node [style=H] (13) at (0.5000001, 0.75) {H};
		\node [style=H] (14) at (1.5, 0.75) {H};
		\node [style=H] (15) at (1.5, -0.7499999) {H};
		\node [style=H] (16) at (0.5000001, -0.7500001) {H};
		\node [style=H] (17) at (-0.5000001, -0.7500001) {H};
		\node [style=H] (18) at (-1.5, -0.7499999) {H};
	\end{pgfonlayer}
	\begin{pgfonlayer}{edgelayer}
		\draw (4.center) to (11);
		\draw (1.center) to (12);
		\draw (2.center) to (13);
		\draw (3.center) to (14);
		\draw (16) to (7.center);
		\draw (15) to (8.center);
		\draw (17) to (6.center);
		\draw (18) to (5.center);
		\draw [bend left=15, looseness=1.00] (18) to (0);
		\draw [bend left=15, looseness=1.00] (17) to (0);
		\draw [bend right, looseness=0.75] (16) to (0);
		\draw [bend right=15, looseness=1.00] (15) to (0);
		\draw [bend right=15, looseness=1.00] (0) to (13);
		\draw [bend right=15, looseness=1.00] (0) to (14);
		\draw [bend left=15, looseness=1.00] (0) to (12);
		\draw [bend left=15, looseness=1.00] (0) to (11);
	\end{pgfonlayer}
\end{tikzpicture}

%% file: tikz/ZXpt_euler_L.tikz
\begin{tikzpicture}
	\begin{pgfonlayer}{nodelayer}
		\node [style=H] (0) at (0, -0) {H};
		\node [style=none] (1) at (0, 0.5) {};
		\node [style=none] (2) at (0, -0.5) {};
	\end{pgfonlayer}
	\begin{pgfonlayer}{edgelayer}
		\draw (2.center) to (0);
		\draw (0) to (1.center);
	\end{pgfonlayer}
\end{tikzpicture}

%% file: tikz/ZXpt_euler_R.tikz
\begin{tikzpicture}
	\begin{pgfonlayer}{nodelayer}
		\node [style=gn] (0) at (0, 1) {$-\frac{\pi}{2}$};
		\node [style=gn] (1) at (0, -1) {$-\frac{\pi}{2}$};
		\node [style=rn] (2) at (0, -0) {$-\frac{\pi}{2}$};
		\node [style=none] (3) at (0, 1.5) {};
		\node [style=none] (4) at (0, -1.5) {};
	\end{pgfonlayer}
	\begin{pgfonlayer}{edgelayer}
		\draw (4.center) to (1);
		\draw (1) to (2);
		\draw (2) to (0);
		\draw (0) to (3.center);
	\end{pgfonlayer}
\end{tikzpicture}

%% file: tikz/ZXpt_D1_L.tikz
\begin{tikzpicture}
	\begin{pgfonlayer}{nodelayer}
		\node [style=rn] (0) at (0, -0.25) {$\quad$};
		\node [style=gn] (1) at (0, 0.25) {$\quad$};
	\end{pgfonlayer}
	\begin{pgfonlayer}{edgelayer}
		\draw (1) to (0);
	\end{pgfonlayer}
\end{tikzpicture}

%% file: tikz/ZXpt_D1_R.tikz
\begin{tikzpicture}
	\begin{pgfonlayer}{nodelayer}
		\node [style=diam] (0) at (0, -0) {};
	\end{pgfonlayer}
\end{tikzpicture}

%% file: tikz/ZXpt_D2_L.tikz
\begin{tikzpicture}
	\begin{pgfonlayer}{nodelayer}
		\node [style=diam] (0) at (-0.25, -0) {};
		\node [style=diam] (1) at (0.25, -0) {};
	\end{pgfonlayer}
\end{tikzpicture}

%% file: tikz/ZXpt_D2_R.tikz
\begin{tikzpicture}
	\begin{pgfonlayer}{nodelayer}
		\node [style=none] (0) at (0, 0.5) {};
		\node [style=none] (1) at (-0.5, -0) {};
		\node [style=none] (2) at (0, -0.5) {};
		\node [style=none] (3) at (0.5, -0) {};
	\end{pgfonlayer}
	\begin{pgfonlayer}{edgelayer}
		\draw [bend left=45, looseness=1.00] (1.center) to (0.center);
		\draw [bend left=45, looseness=1.00] (0.center) to (3.center);
		\draw [bend right=45, looseness=1.00] (2.center) to (3.center);
		\draw [in=-90, out=180, looseness=1.00] (2.center) to (1.center);
	\end{pgfonlayer}
\end{tikzpicture}

%% file: incompleteness.tex
Before we present the proof, let us recall a standard result from quantum
mechanics, namely the \textit{Euler decomposition} of single-qubit gates
\cite{quantum_book}. By this result, any single-qubit unitary gate can be
expressed (up to a global phase) through just three consecutive rotations in
appropriate bases. For the ZX-calculus, this means that there always exist real
angles 
$\alpha_i,\beta_i,\gamma_i, \phi_i$ such that for any ZX diagram $D$ with one
input and one output, we have:
\begin{center} 
  $
  \left \llbracket \quad 
    %
\InputIfFileExists{tikz/euler1.tikz}{}{\input{./figures/tikz/euler1.tikz}}

  \quad \right \rrbracket
  $
  = $
  e^{i\phi_1}
  \left \llbracket \quad 
  %
\InputIfFileExists{tikz/euler2.tikz}{}{\input{./figures/tikz/euler2.tikz}}

  \quad \right \rrbracket
  $ 
  = 
  $
  e^{i\phi_2}
  \left \llbracket \quad 
  %
\InputIfFileExists{tikz/euler3.tikz}{}{\input{./figures/tikz/euler3.tikz}}

  \quad \right \rrbracket
  $ 
\end{center}

We prove the incompleteness of the ZX-calculus by using a similar argument to
that of Duncan and Perdrix in \cite{euler_necessity}, where they show that the
Euler decomposition of the Hadamard gate is not derivable within the ZX-calculus.

In particular, we can define alternative models for the ZX-calculus by setting:

\begin{center}
\input{models.tex}
\end{center}
\begin{align*}
\llbracket \cdot \rrbracket_k := \llbracket \cdot \rrbracket
\text{ , otherwise}
\end{align*}

where $k \in \mathbb{Z}$ and $\llbracket \cdot \rrbracket$ is the standard
interpretation functor for ZX diagrams in Hilbert space. In other words, we
multiply all angles in our diagrams by an integer $k$ and consider the
corresponding interpretation.

These models are sound when $k=4p+1$ for $p \in \mathbb{Z}$. This can be easily
verified by checking that each of the equational rules remains valid under this
interpretation.

Consider the following two ZX diagrams:

\begin{align*}
\begin{tikzpicture}
	\node [style=none] (o)   at (0, 8)   {};
	\node [style=none] (D1)  at (0, 4)   {$D_1 :=$};
	\node [style=none] (i)   at (0, 0)   {};
\end{tikzpicture}
\quad
\input{tikz/LHS.tikz}
\quad\quad
\begin{tikzpicture}
	\node [style=none] (o)   at (0, 8)   {};
	\node [style=none] (D2)  at (0, 4)   { and\quad\quad $D_2 :=$};
	\node [style=none] (i)   at (0, 0)   {};
\end{tikzpicture}
\quad
\input{tikz/RHS.tikz}
\end{align*}
where
\begin{align*}
\alpha &:= - \arccos \left ( \frac 5 {2\sqrt{13}} \right )
  \approx 0.2561 \pi \\
\beta  &:= -2 \arcsin \left ( \frac {\sqrt 3} 4\right ) 
  \approx -0.2851 \pi \\
\phi   &:= \arcsin \left ( \frac {\sqrt 3} 4\right ) - \alpha 
  \approx 0.3987 \pi
\end{align*}
Then, we have 
\begin{align*}
\llbracket D_1 \rrbracket = \llbracket D_2 \rrbracket
\end{align*}
The two scalar factors are introduced so that the equality is exact, otherwise
it would be true up to the global phase $e^{i\phi}$.

Let us assume for contradiction that $D_1$ and $D_2$ are equal under the axioms
of the ZX-calculus, i.e. \\$ZX \vdash D_1 = D_2.$ Since
$\llbracket \cdot \rrbracket_{-3}$ provides a sound model of the calculus, it
must also be the case that $\llbracket D_1 \rrbracket_{-3} = \lambda \llbracket
D_2 \rrbracket_{-3},$ for some $\lambda \in \mathbb{C}$.

However, it is easy to check that this is not true ($\llbracket D_1
\rrbracket_{-3}$ is equal to a scalar times the identity, whereas $\llbracket
D_2 \rrbracket_{-3}$ isn't). Therefore the two diagrams $D_1$ and $D_2$ are not
equal under the axioms of the ZX-calculus, even though they have equal Hilbert
space interpretations. This means the ZX-calculus is incomplete.

%% file: tikz/euler1.tikz
\begin{tikzpicture}
	\begin{pgfonlayer}{nodelayer}
		\node [style=block] (2) at (0, -0) {$D$};
		\node [style=none] (3) at (0, -1.25) {};
		\node [style=none] (4) at (0, 1.25) {};
	\end{pgfonlayer}
	\begin{pgfonlayer}{edgelayer}
		\draw (3) to (2);
		\draw (4) to (2);
	\end{pgfonlayer}
\end{tikzpicture}

%% file: tikz/euler2.tikz
\begin{tikzpicture}
	\begin{pgfonlayer}{nodelayer}
		\node [style=rn] (0) at (0, 0.75) {$\alpha_1$};
		\node [style=gn] (1) at (0, -0) {$\beta_1$};
		\node [style=rn] (2) at (0, -0.75) {$\gamma_1$};
		\node [style=none] (3) at (0, -1.25) {};
		\node [style=none] (4) at (0, 1.25) {};
	\end{pgfonlayer}
	\begin{pgfonlayer}{edgelayer}
		\draw (3.center) to (2);
		\draw (2) to (1);
		\draw (1) to (0);
		\draw (0) to (4.center);
	\end{pgfonlayer}
\end{tikzpicture}

%% file: tikz/euler3.tikz
\begin{tikzpicture}
	\begin{pgfonlayer}{nodelayer}
		\node [style=gn] (0) at (0, 0.75) {$\alpha_2$};
		\node [style=rn] (1) at (0, -0) {$\beta_2$};
		\node [style=gn] (2) at (0, -0.75) {$\gamma_2$};
		\node [style=none] (3) at (0, -1.25) {};
		\node [style=none] (4) at (0, 1.25) {};
	\end{pgfonlayer}
	\begin{pgfonlayer}{edgelayer}
		\draw (3.center) to (2);
		\draw (2) to (1);
		\draw (1) to (0);
		\draw (0) to (4.center);
	\end{pgfonlayer}
\end{tikzpicture}

%% file: models.tex
$
\left \llbracket \quad
\begin{aligned}
\begin{tikzpicture}
	\begin{pgfonlayer}{nodelayer}
		\node [style=gn] (0) at (0, -0) {$\alpha$};
		\node [style=none] (1) at (0.5000003, 1.5) {};
		\node [style=none] (2) at (-0.5000003, 1.5) {};
		\node [style=none] (3) at (0.5000003, -1.5) {};
		\node [style=none] (4) at (-0.5000003, -1.5) {};
		\node [style=none] (5) at (0, 1.25) {...};
		\node [style=none] (6) at (0, -1.25) {...};
	\end{pgfonlayer}
	\begin{pgfonlayer}{edgelayer}
		\draw [bend right=15, looseness=1.00] (2.center) to (0);
		\draw [bend right=15, looseness=1.00] (0) to (1.center);
		\draw [bend right=15, looseness=1.00] (0) to (4.center);
		\draw [bend left=15, looseness=1.00] (0) to (3.center);
	\end{pgfonlayer}
\end{tikzpicture}
\end{aligned}
\mbox{} \quad \right \rrbracket_k :=
$
$
\left \llbracket \quad
\begin{aligned}
\begin{tikzpicture}
	\begin{pgfonlayer}{nodelayer}
		\node [style=gn] (0) at (0, -0) {$k\alpha$};
		\node [style=none] (1) at (0.5000003, 1.5) {};
		\node [style=none] (2) at (-0.5000003, 1.5) {};
		\node [style=none] (3) at (0.5000003, -1.5) {};
		\node [style=none] (4) at (-0.5000003, -1.5) {};
		\node [style=none] (5) at (0, 1.25) {...};
		\node [style=none] (6) at (0, -1.25) {...};
	\end{pgfonlayer}
	\begin{pgfonlayer}{edgelayer}
		\draw [bend right=15, looseness=1.00] (2.center) to (0);
		\draw [bend right=15, looseness=1.00] (0) to (1.center);
		\draw [bend right=15, looseness=1.00] (0) to (4.center);
		\draw [bend left=15, looseness=1.00] (0) to (3.center);
	\end{pgfonlayer}
\end{tikzpicture}\end{aligned}
\mbox{} \quad \right \rrbracket
$
\quad ; \quad
$
\left \llbracket \quad
\begin{aligned}
\begin{tikzpicture}
	\begin{pgfonlayer}{nodelayer}
		\node [style=rn] (0) at (0, -0) {$\alpha$};
		\node [style=none] (1) at (0.5000003, 1.5) {};
		\node [style=none] (2) at (-0.5000003, 1.5) {};
		\node [style=none] (3) at (0.5000003, -1.5) {};
		\node [style=none] (4) at (-0.5000003, -1.5) {};
		\node [style=none] (5) at (0, 1.25) {...};
		\node [style=none] (6) at (0, -1.25) {...};
	\end{pgfonlayer}
	\begin{pgfonlayer}{edgelayer}
		\draw [bend right=15, looseness=1.00] (2.center) to (0);
		\draw [bend right=15, looseness=1.00] (0) to (1.center);
		\draw [bend right=15, looseness=1.00] (0) to (4.center);
		\draw [bend left=15, looseness=1.00] (0) to (3.center);
	\end{pgfonlayer}
\end{tikzpicture}
\end{aligned}
\mbox{} \quad \right \rrbracket_k :=
$
$
\left \llbracket \quad
\begin{aligned}
\begin{tikzpicture}
	\begin{pgfonlayer}{nodelayer}
		\node [style=rn] (0) at (0, -0) {$k\alpha$};
		\node [style=none] (1) at (0.5000003, 1.5) {};
		\node [style=none] (2) at (-0.5000003, 1.5) {};
		\node [style=none] (3) at (0.5000003, -1.5) {};
		\node [style=none] (4) at (-0.5000003, -1.5) {};
		\node [style=none] (5) at (0, 1.25) {...};
		\node [style=none] (6) at (0, -1.25) {...};
	\end{pgfonlayer}
	\begin{pgfonlayer}{edgelayer}
		\draw [bend right=15, looseness=1.00] (2.center) to (0);
		\draw [bend right=15, looseness=1.00] (0) to (1.center);
		\draw [bend right=15, looseness=1.00] (0) to (4.center);
		\draw [bend left=15, looseness=1.00] (0) to (3.center);
	\end{pgfonlayer}
\end{tikzpicture}
\end{aligned}
\mbox{} \quad \right \rrbracket
$

%% file: tikz/LHS.tikz
\begin{tikzpicture}
	\node [style=none] (o)   at (0, 8)   {};
	\node [style=gn]  (g5)  at (0, 7)   {$\frac \pi 3$};
	\node [style=rn]  (r4)  at (0, 5.5)   {$\frac \pi 3$};
	\node [style=gn]  (g3)  at (0, 4)   {$\frac {2\pi} 3$};
	\node [style=rn]  (r2)  at (0, 2.5)   {$\frac \pi 3$};
	\node [style=gn]  (g1)  at (0, 1)   {$\frac \pi 3$};
	\node [style=none] (i)   at (0, 0)   {};
	\node [style=rn] (s1) at (-1,4.5) {$\quad$};
	\node [style=gn] (s2) at (-1,3.5) {$\quad$};

	\draw [-] (i) to (g1);
	\draw [-] (g1) to (r2);
	\draw [-] (r2) to (g3);
	\draw [-] (g3) to (r4);
	\draw [-] (r4) to (g5);
	\draw [-] (g5) to (o);
	\draw [-] (s1) to (s2);
\end{tikzpicture}

%% file: tikz/RHS.tikz
\begin{tikzpicture}
	\node [style=none] (o)   at (0, 8)   {};
	\node [style=gn]  (r4)  at (0, 5.5)   {$\alpha$};
	\node [style=rn]  (g3)  at (0, 4)   {$\beta$};
	\node [style=gn]  (r2)  at (0, 2.5)   {$\alpha$};
	\node [style=none] (i)   at (0, 0)   {};
	\node [style=gn]  (gl) at  (-1,3.5)   {$\phi$};
	\node [style=rn]  (rl) at  (-1,4.5)   {$\pi$};

	\draw [-] (i) to (r2);
	\draw [-] (r2) to (g3);
	\draw [-] (g3) to (r4);
	\draw [-] (r4) to (o);
	\draw [-] (rl) to (gl);
\end{tikzpicture}

%% file: conclusion.tex
The primary contribution of this work is showing that the ZX-calculus is
incomplete for quantum mechanics. A natural question to ask is what additional
rules can be added to the calculus in order to increase its proving power. The
proof that we have used doesn't use any special properties of the presented
diagrams -- it will be straightforward to apply the same proof to another pair
of single-qubit unitary gates where one of them is the Euler decomposition of
the other. To eliminate this class of counter-examples, we believe that a
"color-swap" rule of the form:

\begin{center}
$
\begin{aligned}
\begin{tikzpicture}
	\node [style=none] (o)   at (0, 5.5)   {};
	\node [style=gn]  (r4)  at (0, 4.5)   {$\alpha_1$};
	\node [style=rn]  (g3)  at (0, 3)   {$\beta_1$};
	\node [style=gn]  (r2)  at (0, 1.5)   {$\gamma_1$};
	\node [style=none] (i)   at (0, 0.5)   {};

	\draw [-] (i) to (r2);
	\draw [-] (r2) to (g3);
	\draw [-] (g3) to (r4);
	\draw [-] (r4) to (o);
\end{tikzpicture}
\end{aligned}
$
\quad=\quad
$
\begin{aligned}
\begin{tikzpicture}
	\node [style=none] (o)   at (0, 5.5)   {};
	\node [style=rn]  (r4)  at (0, 4.5)   {$\alpha_2$};
	\node [style=gn]  (g3)  at (0, 3)   {$\beta_2$};
	\node [style=rn]  (r2)  at (0, 1.5)   {$\gamma_2$};
	\node [style=none] (i)   at (0, 0.5)   {};

	\draw [-] (i) to (r2);
	\draw [-] (r2) to (g3);
	\draw [-] (g3) to (r4);
	\draw [-] (r4) to (o);
\end{tikzpicture}
\end{aligned}
$
\end{center}

might be needed. This would require identifying functions $f_1, f_2, f_3$, s.t.
the above rule is valid and
\begin{align*}
\alpha_1 &= f_1 (\alpha_2, \beta_2, \gamma_2) \\
\beta_1  &= f_2 (\alpha_2, \beta_2, \gamma_2) \\
\gamma_1 &= f_3 (\alpha_2, \beta_2, \gamma_2) \\
\end{align*}
In other words, an analytic solution for converting from ZXZ to XZX Euler
decompositions of single-qubit unitary gates is required. 

Whether the addition of such a 'color-swap' would be sufficient to render the ZX-calculus complete is currently unknown. Some simple candidates for further possible non-derivable equalities are presented here:  
\begin{center}
%
\InputIfFileExists{tikz/zxcomp_flowers_L.tikz}{}{\input{./figures/tikz/zxcomp_flowers_L.tikz}}
 = %
\InputIfFileExists{tikz/zxcomp_flowers_R.tikz}{}{\input{./figures/tikz/zxcomp_flowers_R.tikz}}

\end{center}
\begin{center}
%
\InputIfFileExists{tikz/zxcomp_fences_L.tikz}{}{\input{./figures/tikz/zxcomp_fences_L.tikz}}
 = %
\InputIfFileExists{tikz/zxcomp_fences_R.tikz}{}{\input{./figures/tikz/zxcomp_fences_R.tikz}}

\end{center}
We suggest to conduct numerical investigations into the question of whether such non-derivable equalities between complex ZX-calculus diagrams exist.

\section*{Acknowledgements}

We would like to thank Aleks Kissinger, Miriam Backens and Bob Coecke for constructive
discussions. One of the authors also wishes to gratefully acknowledge support
from the EPSRC and the Scatcherd European Scholarship.

%% file: tikz/zxcomp_flowers_L.tikz
\begin{tikzpicture}
	\begin{pgfonlayer}{nodelayer}
		\node [style=rn] (0) at (0, -0.7499999) {$\gamma$};
		\node [style=gn] (1) at (-0.7500001, -0) {$\beta_1$};
		\node [style=gn] (2) at (0.9999999, -0) {$\beta_N$};
		\node [style=none] (3) at (-0.7500001, 0.75) {};
		\node [style=none] (4) at (0.9999999, 0.75) {};
		\node [style=none] (5) at (0.5, 0.5) {...};
		\node [style=none] (6) at (0, 0.75) {};
		\node [style=gn] (7) at (0, -0) {$\beta_2$};
	\end{pgfonlayer}
	\begin{pgfonlayer}{edgelayer}
		\draw (0) to (1);
		\draw (2) to (0);
		\draw (0) to (7);
		\draw (1) to (3.center);
		\draw (7) to (6.center);
		\draw (2) to (4.center);
	\end{pgfonlayer}
\end{tikzpicture}

%% file: tikz/zxcomp_flowers_R.tikz
\begin{tikzpicture}
	\begin{pgfonlayer}{nodelayer}
		\node [style=gn] (0) at (0, -0.7499999) {$\mathfrak{g}$};
		\node [style=rn] (1) at (-0.7500001, -0) {$\mathfrak{b}_1$};
		\node [style=rn] (2) at (0.9999999, -0) {$\mathfrak{b}_N$};
		\node [style=none] (3) at (-0.7500001, 0.75) {};
		\node [style=none] (4) at (0.9999999, 0.75) {};
		\node [style=none] (5) at (0.5, 0.5) {...};
		\node [style=none] (6) at (0, 0.75) {};
		\node [style=rn] (7) at (0, -0) {$\mathfrak{b}_2$};
	\end{pgfonlayer}
	\begin{pgfonlayer}{edgelayer}
		\draw (0) to (1);
		\draw (2) to (0);
		\draw (0) to (7);
		\draw (1) to (3.center);
		\draw (7) to (6.center);
		\draw (2) to (4.center);
	\end{pgfonlayer}
\end{tikzpicture}

%% file: tikz/zxcomp_fences_L.tikz
\begin{tikzpicture}
	\begin{pgfonlayer}{nodelayer}
		\node [style=rn] (0) at (-0.5, -0) {$\beta_1$};
		\node [style=gn] (1) at (0.5, -0) {$\beta_2$};
		\node [style=rn] (2) at (0.5, 0.75) {$\alpha_2$};
		\node [style=rn] (3) at (0.5, -0.75) {$\gamma_2$};
		\node [style=gn] (4) at (-0.5, -0.75) {$\gamma_1$};
		\node [style=gn] (5) at (-0.5, 0.75) {$\alpha_1$};
		\node [style=none] (6) at (-0.5, 1.25) {};
		\node [style=none] (7) at (0.5, 1.25) {};
		\node [style=none] (8) at (-0.5, -1.25) {};
		\node [style=none] (9) at (0.5, -1.25) {};
	\end{pgfonlayer}
	\begin{pgfonlayer}{edgelayer}
		\draw (0) to (1);
		\draw (0) to (5);
		\draw (0) to (4);
		\draw (1) to (3);
		\draw (2) to (1);
		\draw (5) to (6.center);
		\draw (2) to (7.center);
		\draw (8.center) to (4);
		\draw (3) to (9.center);
	\end{pgfonlayer}
\end{tikzpicture}

%% file: tikz/zxcomp_fences_R.tikz
\begin{tikzpicture}
	\begin{pgfonlayer}{nodelayer}
		\node [style=rn] (0) at (0.5, -0) {$\mathfrak{b}_2$};
		\node [style=gn] (1) at (-0.5, 0) {$\mathfrak{b}_1$};
		\node [style=rn] (2) at (-0.5, -0.75) {$\mathfrak{c}_1$};
		\node [style=rn] (3) at (-0.5, 0.75) {$\mathfrak{a}_1$};
		\node [style=gn] (4) at (0.5, 0.75) {$\mathfrak{a}_2$};
		\node [style=gn] (5) at (0.5, -0.75) {$\mathfrak{c}_2$};
		\node [style=none] (6) at (0.5, -1.25) {};
		\node [style=none] (7) at (-0.5, -1.25) {};
		\node [style=none] (8) at (0.5, 1.25) {};
		\node [style=none] (9) at (-0.5, 1.25) {};
	\end{pgfonlayer}
	\begin{pgfonlayer}{edgelayer}
		\draw (0) to (1);
		\draw (0) to (5);
		\draw (0) to (4);
		\draw (1) to (3);
		\draw (2) to (1);
		\draw (5) to (6.center);
		\draw (2) to (7.center);
		\draw (8.center) to (4);
		\draw (3) to (9.center);
	\end{pgfonlayer}
\end{tikzpicture}